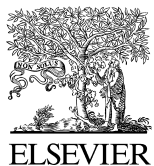

# Decays of the $f_0(1370)$ scalar glueball candidate [*]

Ugo Gastaldi[**]

*INFN-Ferrara, Via Saragat 1, 44122 Ferrara, Italy*

**Abstract**

For a long time doubts have existed on the existence of the $f_0(1370)$ meson as an individual object distinct and separated from the $\sigma$ meson. Decays into $\pi^+\pi^-$ of the $f_0(1370)$ are the main source of an isolated structure localized between 1.2 and 1.5 GeV in the $\pi^+\pi^-$ mass spectrum measured in pp Central Exclusive Production (CEP) at $\sqrt{s}$=200 GeV at very low four momentum transfer squared |t|. These data confirm in the $\pi^+\pi^-$ decay channel the existence of the $f_0(1370)$ as an isolated well identified structure that was previously observed in $K^+K^-$, $K_sK_s$, $4\pi^0$, $2\pi^0\pi^+\pi^-$ and $\pi^+\pi^-\pi^+\pi^-$ decays measured in pbar annihilations at rest. The decay branching ratios of $f_0(1370)$ into $\sigma\sigma$, $\rho\rho$, $\pi\pi$, KKbar, $\eta\eta$ relative to $\pi\pi$ decays obtained in analysis of data of pbar annihilations at rest which treat separately $f_0(1370)$ and $\sigma$ are respectively 5.6, 3, 1, 1, 0.02. The decay and production properties of $f_0(1370)$ point to a large gg content. CEP interactions at very high energies favour production of $0^{++}$ and $2^{++}$ mesons. Selection of events with low |t| at both proton vertices suppresses $2^{++}$ structures. LHC runs dedicated to pp CEP measurements at low |t| could then provide a unique clean source of all the low energy scalars, cross check the pbar annihilation results and make it clear if and where scalar gluonium is resident and the nature (composition in terms of qqbar, qqqbarqbar, qqbar-qqbar and gg) of $f_0(500)$, $f_0(980)$, $f_0(1370)$, $f_0(1500)$ and $f_0(1710)$.

*Keywords*: glueballs, scalar mesons, pbar annihilations, central exclusive production

## 1. Introduction

The possibility of the existence of glueballs is a basic qualitative prediction of quantum chromodynamics (QCD)[1,2]. The lowest lying glueball states are expected to have the same $0^{++}$ $J^{PC}$ quantum numbers as the vacuum. While for the $0^-$ pseudoscalar, $1^-$ vector and $2^{++}$ tensor meson nonets there are two observed isoscalar partners for two places in each nonet (respectively $\eta$ and $\eta$', $\omega$ and $\varphi$, $f_2(1270)$ and $f'_2(1525)$, there are 4 or 5 observed isoscalar mesons ($\sigma$, $f_0(980)$, $f_0(1370)$, $f_0(1500)$ and $f_0(1710)$ for the two places of the isoscalar members of the $0^{++}$ ground state nonet. The candidates for the two places are 4 or 5 depending whether $\sigma$ and $f_0(1370)$ are considered as distinct separated objects or they are part of a single continuum. The existence of the $\sigma$ meson is considered established since some years (see [3] for a review) and $\sigma$ is currently called $f_0(500)$. Doubts instead concern the existence of the $f_0(1370)$ as an individual isolated structure (see [4] for a review). Concerning the nature of the $\sigma$ and $f_0(980)$ $0^{++}$ mesons, various scenarios envisaged in the literature include (because of mixing) the possibility of a qqbar, qqqbarqbar, qqbar-qqbar and gg content in their wave functions[3-7]. If the $0^{++}$ isoscalars are 5 it is more difficult to exclude the hypothesis of the presence (or even dominance) of a gg content in some of them. Production in glue rich processes (central production mediated by double pomeron exchange, pbar annihilations, J/$\psi$ and $\psi$' radiative decays, heavy meson decays) and absence in $\gamma\gamma$ production and meson exchange give criteria to characterize the gg content of the scalars [8]. Relative decay branching ratios into $\sigma\sigma$, $\rho\rho$, $\pi\pi$, KKbar, $\eta\eta$ for the three heavier scalars, and also into $\eta\eta$' for $f_0(1500)$ and $f_0(1710)$, give other selection criteria to identify their gg content [8-11]. Currently there is no consensus concerning the experimental observation of a scalar glueball nor on the possible gg content in the scalars experimentally observed (for reviews see [4,12-17]).

An isolated $\pi^+\pi^-$ peak between 1.2 and 1.5 GeV practically without background and well separated from the $f_0(980)$ signal by a valley nearly devoid of events is present in pp Central Exclusive Production (CEP) data of the STAR experiment [18]

---

and has been interpreted as a clear manifestation that the $f_0(1370)$ meson is an isolated structure of limited width [19,20] (see fig 1 from ref [18]). This interpretation is based on STAR data [18, 21-25], on the scaling laws of CEP [8,26] and on results of the AFS experiment [27,28] at the CERN ISR.

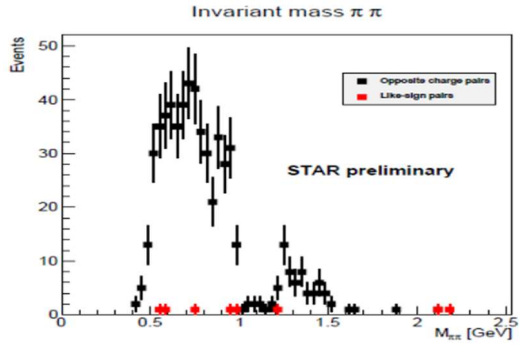

Figure 1: $\pi^+\pi^-$ mass plot of STAR pp CEP raw data of the 200 GeV 2009 run with $0.003<|t_1|, |t_2|<0.03$ GeV$^2$ kinematic coverage of both scattered protons (Fig. from ref.[18]).

The STAR data have been discussed extensively in refs [19,20,29]. In this written version of my contribution to QCD 18 the focus is on data of pbar annihilation at rest which show qualitatively that $f_0(1370)$ is an isolated well identifiable structure and on quantitative results of data analysis which introduced the hypothesis (validated by the STAR data) that $f_0(1370)$ could be an isolated structure.

In section 2 we recall data where the $f_0(1370)$ is present as an isolated clearly visible structure in $K^+K^-$, $K_sK_s$, $\pi^+\pi^-\pi^+\pi^-$, $2\pi^0\pi^+\pi^-$ and $4\pi^0$ decays measured in pbar annihilations at rest. Some of these data have been ignored in the PDG compilations of the last 18 years concerning $f_0(1370)$ and $f_0(1500)$, some are used in data averages to give mass, width and decay branching ratios of $f_0(1500)$, and are discarded in the $f_0(1370)$ section. Besides establishing –together with the STAR data- the existence of the $f_0(1370)$ as an isolated structure, the antiproton data are essential to extract the production rates and decay branching ratios of both $f_0(1370)$ and $f_0(1500)$, in order to identify their gg content.

In section 3 are discussed production and decay branching ratios of $f_0(1370)$ and $f_0(1500)$ in pbar annihilations at rest in liquid $H_2$ targets. The orders of magnitude of the decay branching ratios of $f_0(1370)$ into $\sigma\sigma$, $\rho\rho$, $\pi\pi$, $KKbar$, $\eta\eta$ relative to $\pi\pi$ decays are respectively 5.6, 3, 1, 1, 0.02. These data point at a large gg content of the $f_0(1370)$ meson and need a definitive confirmation by CEP measurements at LHC, also because there are conflicting results of the WA102 experiment [30-32] based on low energy pp CEP data.

Section 4 discusses briefly presence of $f_0(1370)$ in $J/\psi$ and $\psi'$ radiative decays and heavy meson decays and absence in $\gamma\gamma$ production and meson exchange. Section 5 gives conclusions, stresses the importance of the assessment of the properties of the $f_0(1370)$ for the observation of a scalar glueball and highlights the substantial improvements achievable by pp CEP experiments at LHC in the medium and longer term for the spectroscopy of all low energy scalars.

## 2. Direct evidences of the existence of the $f_0(1370)$ meson as an isolated structure

The $3\pi^0$ Dalitz plot of pbar annihilations at rest in liquid $H_2$ features one narrow uniformly populated $2\pi^0$ band which has prompted the identification of the $f_0(1500)$ scalar meson (see fig. 1 in ref [33]) in spite of the absence of an isolated $\pi^0\pi^0$ peak at 1500 MeV in the $\pi^0\pi^0$ mass plot. No direct visual evidence of the $f_0(1370)$ is instead present in this $3\pi^0$ Dalitz plot.

Decays of the $f_0(1500)$ into $2\pi^0$ in npbar $\to \pi^-2\pi^0$ annihilations at rest in liquid $D_2$ are visible in the $\pi^-2\pi^0$ Dalitz plot (see fig. 1 in ref.. [34]), but are obscured in the associated $2\pi^0$ mass plot by the reflections of the crossing of the two $\rho^-$ bands. Decays of the $f_0(1370)$ into $2\pi^0$ in npbar $\to \pi^-2\pi^0$ annihilations at rest in liquid $D_2$ may be suggested by a slight distortion of the $f_2(1270)$ band, but are masked by the preponderant signal of the $f_2(1270)$ in the associated $2\pi^0$ mass plot.

In summary a convincing visual signal of $\pi\pi$ decays of $f_0(1500)$ is present only in $2\pi^0$ decays while it is not present in $\pi^+\pi^-$ decays in pbar annihilations at rest in liquid $H_2$ and $D_2$ targets.

Inspections of $\pi^+\pi^-\pi^0$, $\pi\pi^0\pi^0$ and $3\pi^0$ Dalitz plots do not suggest the presence of the $f_0(1370)$ structure between the $f_2(1270)$ and the $f_0(1500)$ (see e.g. fig. 6 in ref. [35]).

Both the $f_0(1370)$ and the $f_0(1500)$ signals are visible in the $\pi^0\eta\eta$ Dalitz plot and in the associated $\eta\eta$ mass plot of ppbar $\to \pi^0\eta\eta$ annihilations at rest in liquid $H_2$ (see fig. 1 in ref. [36]). In ppbar $\to \pi^0\eta\eta$ annihilations at rest in gaseous $H_2$ at 12 atm both the $f_0(1370)$ and the $f_0(1500)$ signal are visible in the $\pi^0\eta\eta$ Dalitz plot and in the associated $\eta\eta$ mass plot (see fig. 1 in ref [37]), but the $f_0(1370)$ signal appears weaker while the $f_0(1500)$ looks stronger compared to the signals in liquid $H_2$. The ratios of ppbar $J^{PC}$ atomic initial states which contribute to annihilations change when the $H_2$ target density changes [38]. The fraction of P-wave initial states of annihilation increases at lower target densities. Preferential $f_0(1370)$ production from S-wave initial states explains the difference between the ppbar $\to \pi^0\eta\eta$ data at the two target densities. The data and analysis of pbar annihilation at rest discussed above are used and quoted in the PDG compilations and in most glueball reviews. They provide values for the relative decay branching ratios into $\pi\pi$ and $\eta\eta$ of $f_0(1370)$ and $f_0(1500)$ discussed in the next section. The situation is different for what concerns the KKbar and 4 pion decays of $f_0(1370)$ and $f_0(1500)$. Evidence for the existence of the $f_0(1370)$ scalar as a structure of limited width (200-300 MeV) and mass centered near 1370 MeV is present since long in data of pbar annihilations which are not quoted in the PDG compilations or which are quoted and used in the $f_0(1500)$ section of the PDG compilations of the last two decades to give average values for mass and width of $f_0(1500)$, but are discarded in the $f_0(1370)$ section.

Decays of $f_0(1370)$ into $K^+K^-$ and into $K_sK_s$ are directly observable in $\pi^0K^+K^-$, $\pi^-K_sK_s$ and $\pi^0K_sK_s$ Dalitz plots and in the respective $K^+K^-$ and $K_sK_s$ mass plots of pp annihilations at rest in liquid $H_2$ and np annihilations at rest in liquid $D_2$ targets (see fig. 2 from ref. [39]). $K^+K^-$ decays of $f_0(1370)$ are the main source of the diagonal $K^+K^-$ band comprised between the $K^{*+}$ and $K^{*-}$ bands in the $\pi^0K^+K^-$ Dalitz plot of the top left plate of fig.2 and of the peak at 1.4 GeV in the $K^+K^-$ mass plot of the top right plate of fig.2. Decays into $K^+K^-$ of the $f_0(1500)$ are not easily visible in ppbar annihilations in liquid $H_2$ targets [40-42] (neither in the $\pi^0K^+K^-$ Dalitz plot, nor in the $K^+K^-$ mass plot) because the $f_0(1500)$ signal is weak in liquid $H_2$ targets, since it is only partially produced from S-wave initial atomic states (while the $f_0(1370)$ signal is dominantly produced from $0^{-+}$ S-wave initial atomic states [42]), and S-wave annihilations dominate in liquid $H_2$ targets (while P-wave annihilations dominate in low density $H_2$ targets). Moreover the weak $f_0(1500)$ signal may be shadowed by the nearly overlapping $f_2'(1525)$ signal.

Destructive interference of the $f_0(1500)$ amplitude with the $f_0(1370)$ amplitude generates the deep narrow flat valley at $M(K_sK_s)=1.5$ GeV in the $\pi^0K_sK_s$ Dalitz plot and in the $K_sK_s$ mass plot obtained [43] using CERN [44,45] and BNL [46] bubble chamber data (see the bottom left and right plates in fig. 2). This $K_sK_s$ valley in the $\pi^0K_sK_s$ bubble chamber Dalitz plot and in the $K_sK_s$ mass plot represents probably the best evidence of KKbar decays of $f_0(1500)$. The same effect is visible in the high statistics $\pi^0K_lK_l$ Dalitz plot and in the associated $K_lK_l$ mass plot produced by Crystal Barrel [47] (see figs.1 and 2 in [47]), where the valley is half as deep because of background and less mass resolution than in the $\pi^0K_sK_s$ Dalitz plot (notice that ref.[47] uses the frequency of ppbar $\rightarrow \pi^0K_sK_s$ annihilations in liquid $H_2$ as frequency of the the ppbar $\rightarrow \pi^0K_lK_l$ annihilations).

In the $\pi^-K_sK_s$ Dalitz plot produced [48] using CERN [49] and BNL [50] bubble chamber data (see the central left plate in fig. 2) the effect of the interference of the $f_0(1370)$ band with the $K^*$ bands is clearly noticeable. The $K_sK_s$ decays of $f_0(1370)$ are the main source of the peak at about 1.4 GeV in the $K_sK_s$ mass plot of the central right plate of fig.2. These data represent probably the best direct evidence of KKbar $f_0(1370)$ decays. A direct quantitative comparison of $\pi^-K_sK_s$ and $\pi^-2\pi^0$ annihilation data in liquid $D_2$ would help providing relative decay branching ratios of KKbar and $\pi\pi$ decays of $f_0(1370)$, but it is not available.

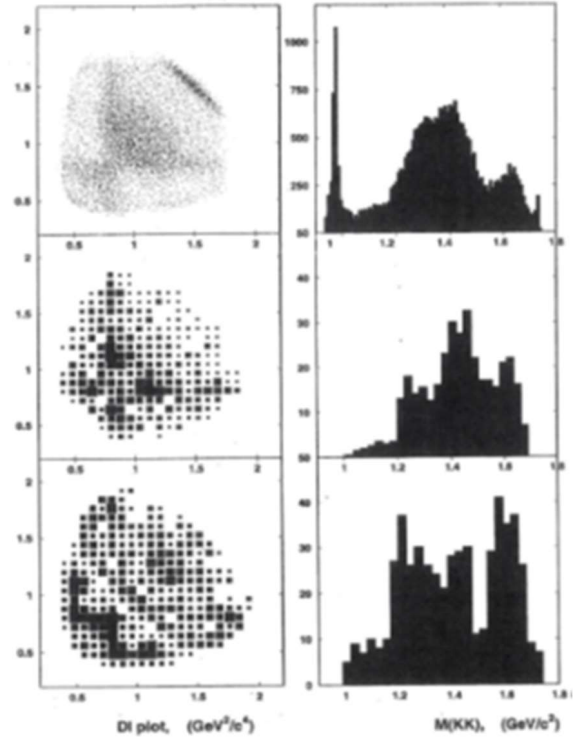

Figure 2: $\pi^0K^+K^-$, $\pi^-K_sK_s$, $\pi^0K_sK_s$ Dalitz plots (top, middle and bottom left plates) and respective $K^+K^-$ and $K_sK_s$ mass plots (right plates) of pbar annihilations at rest in liquid $H_2$ and $D_2$ targets (picture from ref. [39])

Data of ppbar annihilations at rest into three final states ($\pi^+\pi^-\pi^0$, $K^+K^-\pi^0$ and $K_sK^-\pi^+/K_sK^-\pi^+$) at 3 densities of the $H_2$ target ( liquid $H_2$, NTP gaseous $H_2$ and 5 mbar NT gaseous $H_2$) have been collected and analyzed by the Obelix experiment at LEAR in a coupled channel analysis [42,51]. The spectra at the 3 target densities have markedly different shapes because the relative production of intermediate resonances ($\pi\pi$, KKbar, $\pi K$) depends dramatically on the $J^{PC}$ initial atomic states of annihilation, and the $J^{PC}$ fractions of annihilation depend substantially on the target density [38]. The $K^+K^-\pi^0$ data in liquid $H_2$ constrain the width of the $f_0(1370)$ so that the hypothesis of a broad $f_0(1370)$ extending below 1 GeV is naturally discarded, and the $\pi^+\pi^-\pi^0$ data are fit with sensible priors for the $f_0(1370)$ and the $f_0(1500)$ width and mass. The $f_0(1370)$ is essentially produced only from $J^{PC} = 0^{-+}$ S-wave initial states while for $f_0(1500)$ there is a relevant production from $J^{PC} = 1^{++}$ P-wave initial states [42] (this feature may explain why $f_0(1500)$ and not $f_0(1370)$ are observed in ppbar annihilations in flight at 900 and at 1640 MeV/c [52] ). The coupled channel analysis shows that the ratio between the KKbar and $\pi\pi$ couplings of $f_0(1370)$ is of the order of 1 [42,51].

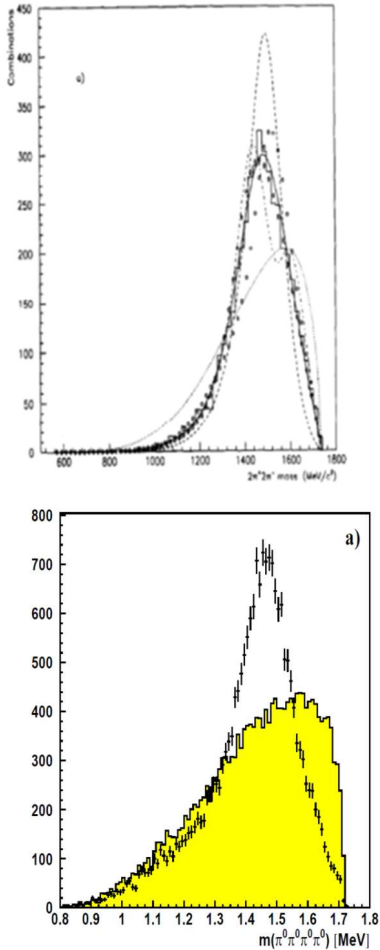

Figure 3: $\pi^+\pi^-\pi^+\pi^-$ invariant mass in npbar → $\pi^+\pi^-\pi^+\pi^-$ annihilations in liquid $D_2$; the histogram shows data, the dotted line shows phase space (top plate, from ref.[54]). $4\pi^0$ invariant mass in npbar → $\pi^-\,4\pi^0$ annihilations at rest in liquid $D_2$; the peak in data points with errors is mainly due to $f_0(1370)$ decays to $4\pi^0$, the colored histogram shows phase space (bottom plate, from [57]).

Decays of $f_0(1370)$ into $\pi^+\pi^-\pi^+\pi^-$ and into $4\pi^0$ measured in $\pi^-\pi^+\pi^-\pi^+\pi^-$ [53,54] and into $\pi^-\,4\pi^0$ [55-57] pbar annihilations at rest in liquid $D_2$ are the dominant feature (well distinct from phase space) in the respective mass plots (see fig.3 top plate from ref. [54] and fig. 3 bottom plate from ref.[57]). The $4\pi^0$ decays are particularly interesting because they do not suffer from combinatorial background and cannot be the result of $\rho\rho$ decays. These data plus five pion annihilation data in liquid $H_2$ [58,59] show that the $f_0(1370)$ scalar decays dominantly into $\sigma\sigma$ (with a frequency twice that of the $\rho\rho$ decays and about 6 times larger than $\pi\pi$ decays) and that $\sigma\sigma$ and $\rho\rho$ decays of $f_0(1370)$ are 10 times more frequent than respectively $\sigma\sigma$ decays and $\rho\rho$ decays of $f_0(1500)$ [57,59].

## 3. Production and decay branching ratios of $f_0(1370)$ and $f_0(1500)$ in pbar annihilations at rest in liquid $H_2$

The $f_0(1370)$ and $f_0(1500)$ scalars overlap appreciably. Their amplitudes interfere. Their decay branching ratios are determined nearly always in the same reactions. We discuss therefore in parallel their production times decays in pbar annihilations at rest. Annihilations of antiprotons at rest have been measured mostly by stopping antiprotons in liquid $H_2$ targets. A major limit in the analysis of data collected only in liquid targets is that the percentage of events produced from annihilations from the S and the P-wave initial states which can contribute to a given channel are not known. The spectra (e.g. Dalitz plots) resulting from different $J^{PC}$ initial states add incoherently. The angular distributions of resonances produced from different $J^{PC}$ initial states are generally different. The angular distribution of a given resonance in an experimental Dalitz plot results from the sum of the different angular distributions of the contributions from the different $J^{PC}$ sources with weights not known a priori. This generates ambiguities and makes it difficult to identify and distinguish resonances with similar energies which contribute to a given final state. S-wave dominance has been generally assumed for annihilations in liquid $H_2$. Coupled channel analysis has been exploited when possible. It restricts the freedom of the fits by imposing the same width in different decay channels of the same resonance. The evolution of the analysis of the various channels of annihilation in liquid $H_2$ passed typically through steps progressing from assumption of pure S-wave annihilation to acceptance of the presence of a significant fraction of P-wave annihilations, to analysis of data collected at different target densities which pin down the S and P-wave contributions and - concerning the $f_0(1370)$- to the introduction in the fits of the hypothesis that the $f_0(1370)$ could have been a resonance of limited width distinct from $f_0(980)$, from $\sigma$ and from a non-resonant S-wave continuum. In the coupled channel analysis of ref. [42,51] events collected at 3 different $H_2$ target densities have permitted to determine at each target density the fractional contributions of the different $J^{PC}$ sources and, for each $J^{PC}$ source, the fractional contribution of each resonance found in the fit. Crystal Barrel high statistics data of $3\pi^0$, $2\pi^0\eta$ and $\pi^0\eta\eta$ annihilations in liquid $H_2$ have undergone two coupled channel analysis [60,61]. In [61] $f_0(1370)$ was treated separately from $f_0(980)$.

We discuss in the following for several annihilation channels the values of the product of the production fraction of the channel in liquid $H_2$ times the decay branching ratio of the $f_0(1370)$ and $f_0(1500)$. The values present in the literature have evolved with time depending on the ambiguities in the treatment of P-wave annihilations, on the assumptions on $f_0(1370)$ width, on the treatment of interferences and on hypothesis on presence, energy and width of isobars in the energy region of the two scalars. Many results have still an uncertainty at the 30% level. However the picture emerging for the ratios of decay branching ratios of the two scalars has evolved and seems now qualitatively clear. In the following we recall the frequencies of several annihilation channels in liquid $H_2$ and $D_2$, (which are measured by counting the number of events and do not depend on assumptions) and the product of frequencies of Production times Decay Branching Ratio (PxDBR) of $f_0(1370)$ and $f_0(1500)$. We quote 2 digits at most for PxDBF, since the uncertainties are likely to be at the 10-30% level:

$\pi^0\eta\eta$ channel frequency $(20 \pm 4)\ 10^{-4}$ [36]

    PxDBR $f_0(1370) \to 2\eta$     $10\ 10^{-4}$ [60]     $0.4\ 10^{-4}$ [61]

    PxDBR $f_0(1500) \to 2\eta$     $6\ 10^{-4}$ [60]     $1.9\ 10^{-4}$ [61]

$3\pi^0$ channel frequency     $(62 \pm 10)\ 10^{-4}$     [33].

    PxDBR $f_0(1370) \to 2\pi^0$     $35\ 10^{-4}$ [60]     $6.4\ 10^{-4}$ [61]

    PxDBR $f_0(1500) \to 2\pi^0$     $13\ 10^{-4}$ [60]     $8.2\ 10^{-4}$ [61]

$\pi^0\pi^+\pi^-$ channel frequency    $(536 \pm 37)\ 10^{-4}$ [62,42]

    PxDBR $f_0(1370) \to \pi^+\pi^-$           $19\ 10^{-4}$ [42]

    PxDBR $f_0(1500) \to \pi^+\pi^-$           $23\ 10^{-4}$ [42]

$K^+K^-\pi^0$ channel frequency    $(23.7 \pm 0.2)\ 10^{-4}$ [40-42]

    PxDBR $f_0(1370) \to K^+K^-$     $6.5\ 10^{-4}$ [41]     $14\ 10^{-4}$ [42]

    PxDBR $f_0(1500) \to K^+K^-$     $0.5\ 10^{-4}$ [41]     $4\ 10^{-4}$ [42]

$\pi^0 K_s K_s$ channel frequency    $(7.5 \pm 0.3)\ 10^{-4}$ [44-46].

    PxDBF $f_0(1370) \to K_s K_s$     $(0.7 \leftrightarrow 5.5)\ 10^{-4}$ [43]

    PxDBF $f_0(1500) \to K_s K_s$     $(2.0 \leftrightarrow 3.8)\ 10^{-4}$ [43]

$\pi^0 K_l K_l$ channel frequency    $(7.5 \pm 0.3)\ 10^{-4}$ [44-46].

    PxDBR $f_0(1370) \to K_l K_l$     $(1.7 \leftrightarrow 4.7)\ 10^{-4}$ [47]

    PxDBR $f_0(1500) \to K_l K_l$     $(1.1 \leftrightarrow 1.9)\ 10^{-4}$ [47]

$\pi^-4\pi^0$ channel frequency    $(67 \pm 10)\ 10^{-4}$ [57]

    PxDBR $f_0(1370) \to \sigma\sigma$:        $46\ 10^{-4}$ [57,59]

    PxDBR $f_0(1500) \to \sigma\sigma$:        $4.5\ 10^{-4}$ [57,59]

$5\pi^0$ channel frequency:    $(71 \pm 10)\ 10^{-4}$ [58]

    PxDBR $f_0(1370) \to \sigma\sigma$:     $28\ 10^{-4}$ [57,59]

    PxDBR $f_0(1500) \to \sigma\sigma$:     $2.8\ 10^{-4}$ [57,59]

$\pi^-2\pi^0\pi^+\pi^-$ channel frequency: $(315 \pm 47)\ 10^{-4}$ [59]
(after removal of events with $\eta$ or $\omega$ decaying into $\pi^0\pi^+\pi^-$)

    PxDBR $f_0(1370) \to \sigma\sigma$ :    $50\ 10^{-4}$ [57,59]

    PxDBR $f_0(1370) \to \rho^+\rho^-$:    $60\ 10^{-4}$ [57,59]

    PxDBR $f_0(1500) \to \sigma\sigma$:    $4.7\ 10^{-4}$ [57,59]

    PxDBR $f_0(1500) \to \rho^+\rho^-$:    $5.4\ 10^{-4}$ [57,59]

Crystal Barrel has also identified the $\pi^*\pi$ and $a_1\pi$ decay channels of $f_0(1370)$ and $f_0(1500)$ in $\pi^-2\pi^0\pi^+\pi^-$ annihilations. By using isospin Clebsch-Gordan coefficients and the ensemble of the 4 pion data in liquid $H_2$ and $D_2$ the production times decay branching ratios in liquid $H_2$ for the $\sigma\sigma$, $\rho\rho$, $\pi^*\pi$, $a_1\pi$ decays have been given in ref [57] and are reproduced below. For the decays of $f_0(1370)$ and $f_0(1500)$ into $\pi\pi$, $\eta\eta$, $\eta\eta'$ two pseudoscalar mesons the more reliable values are in ref [61] and are reproduced below. All these values suffer from the absence of data with lower density $H_2$ targets, which enable the measurement of the contributions from annihilations in P-waves in liquid $H_2$, but result from analysis which identify a $f_0(1370)$ of limited width consistent with the STAR $\pi^+\pi^-$ data.

The values for KKbar decays of $f_0(1500)$ in refs. [41] and [47] are not compatible (and both depend on assumptions). The values of ref. [43] cover a too large window since a direct measurements of the P-wave contributions is missing. For the decays of $f_0(1370)$ and $f_0(1500)$ into KKbar pairs we use the results in ref. [42], which takes experimentally into account P-wave annihilations in liquid $H_2$. The ratios between the $2\pi^0$ and the $\pi^+\pi^-$ decays in [61] and [42] are consistent within 40%. We scale then the $\pi\pi$ and KKbar values derived from [42] according to the ratios between the $\pi\pi$ PxDBR of refs.[42] and [61] for a general comparison.

In summary, by using the values of refs. [57,61] and [42] discussed above, the product of the production times decay branching ratios of $f_0(1370)$ and $f_0(1500)$ into $\sigma\sigma$, $\rho\rho$, $\pi^*\pi$, $a_1\pi$, $\pi\pi$, KKbar, $\eta\eta$, $\eta\eta'$ are (in units of $10^{-4}$) :

   107, 55, 37, 12.5, 19, 19, 0.4, 0   for $f_0(1370)$

   10.5, 5.0, 20, 4.8, 25, 5.6, 1.9, 1.6 for $f_0(1500)$.

The ratios between decay branching ratios from pbar annihilations data at rest, are:

   $f_0(1370)$:   $\sigma\sigma/\rho\rho \approx 2$    $\sigma\sigma/\pi\pi \approx 5.6$    KKbar/$\pi\pi \approx 1$

   $f_0(1500)$ :   $\sigma\sigma/\rho\rho \approx 2$    $\sigma\sigma/\pi\pi \approx 0.4$    KKbar/$\pi\pi \approx 0.2$

Both $\sigma\sigma$ and $\rho\rho$ decays of $f_0(1370)$ and are 10 times more frequent than $\sigma\sigma$ and $\rho\rho$ decays of $f_0(1500)$. The dominance of σσ decays of $f_0(1370)$ is in agreement with the results in [54].

Fig. 4 shows the ratios between the PxDBR of the various decay channels of $f_0(1370)$ and $f_0(1500)$ and the PxDBR of the ππ decay channel. These ratios are proportional to the areas of the 8 sectors of the figure. Since the PxDBRs of the ππ decay channel of the 2 scalars are of the same order, the total area of each figure is proportional to the total production of the scalar in pbar annihilations at rest in liquid $H_2$.

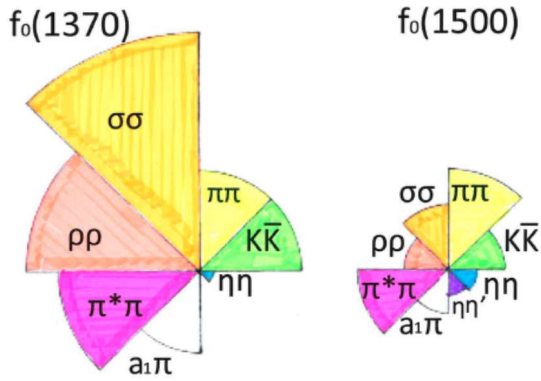

Figure 4: ratios between the decay branching ratio in the 8 identified decay channels of the $f_0(1370)$ and $f_0(1500)$ and their ππ decay branching ratio.

The scenario which emerges for $f_0(1370)$ is compatible with the expectation for a glueball to decay with equal rates into ππ and KKbar deriving from equal couplings of gluons to nnbar and ssbar quarks [1,8] and expectations from QCD sum rules and low energy theorems for a scalar glueball to decay dominantly into σσ [63].

The scenario which emerges for $f_0(1500)$ is far from that of a scalar with a dominant gg component in view of the low KKbar/ππ, σσ/ππ and ηη'/ππ relative branching ratios. Also $f_0(1500)$ seems to have much less ssbar than nnbar content than $f_0(1370)$.

The ensemble of the relative decay branching rates of $f_0(1370)$ and $f_0(1500)$ given above changes substantially the experimental input for all the mixing schemes quoted in ref [17]. There is however the caveat that in the analysis of WA102 pp CEP data at 29 GeV the ρρ decay branching ratio of $f_0(1370)$ is found to be more than 4 times larger than the σσ decay branching ratio [30] and the ratios of the decay branching ratios of $f_0(1370)$ into ππ, KKbar and ηη to the decay branching ratios of $f_0(1370)$ into ππ are of the order of 1, 0,5 and 0,2 [31,32].

## 4. Other $f_0(1370)$ production channels

We comment briefly on presence and/or absence of $f_0(1370)$ in other production channels relevant for glueball searches.

Clear signals of $f_0(1370)$ decays are visible in CLEO spectra of J/ψ and ψ' radiative decays to pairs of pseudoscalar mesons $π^+π^-$, $2π^0$, $K^+K^-$, $K_sK_s$ (see fig 9 in ref. [64]). ππ decays of $f_0(1370)$ appear as an enhancement on the right of a dominant $f_2(1270)$ peak at masses less than 1.5 GeV in $π^+π^-$ and $2π^0$ spectra, and as a signal on the left of of a dominant $f_2'(1525)$ signal in $K^+K^-$ and $K_sK_s$ spectra. It is hard to identify $f_0(1500)$ in ππ spectra in the valley between the $f_2(1270)$ and the $f_0(1710)$ signals, and in the KKbar spectra because of the dominant presence of the $f_2'(1525)$. If one takes the decay br in ππ of $f_0(1500)$ from table 3 in ref [64] as an upper limit for the ππ decay branching ratio of $f_0(1370)$ (which is not found by the fit) and compares it to the decay branching ratio of $f_0(1370)$ to KKbar, the ratio KKbar/ππ for $f_0(1370)$ exceeds 3, while for $f_0(1710)$ the KKbar/ππ ratio is about 3.

The BES spectra of J/ψ radiative decays to $π^+π^-$, $2π^0$, $K^+K^-$, $K_sK_s$ (see figs 2 and 3 in ref. [65] and fig.1 in ref. [66]) display similar topological features to the CLOE ones.

The $f_0(1370)$ signal is clearly noticeable in the very high statistics spectrum of J/ψ radiative decays to $2π^0$ measured by BESIII ( see fig. 1 in ref. [67]). Interference of the $f_0(1500)$ amplitude is likely to be necessary to explain the $0^{++}$ spectrum of fig. 2 in ref [67].

In $B^0_s$bar→J/ψ$π^+π^-$ decays measured by LHCb [68,69] the $π^+π^-$ spectrum features a dominant $f_0(980)$ signal followed by a box of $π^+π^-$ in S-wave that drops down near 1.5 GeV. The rapid drop is likely to be due to the $f_0(1370)$ with a negative interference of $f_0(1500)$ (see figs. 12 and 17 of ref. [68] and figs.15 and 20 in ref. [69]).

In $B^0_s$bar →J/ψ$K^+K^-$ decays measured by LHCb [70,71] the $K^+K^-$ spectrum is dominated by large φ and $f_2'(1525)$ signals whose tails may submerge a possible $f_0(1370)$ signal (see figs.17 in ref [70] and fig. 7 in ref. [71].

In $B^0_s$bar →J/ψ $π^+π^-π^+π^-$ decays measured by LHCb [72] the 4 pion spectrum features a narrow signal due to $f_1(1285)$ and an asymmetric peak occurring before 1.5 GeV (see fig 2 in ref[72], which might be due to the interfering amplitudes of $f_0(1370)$ and $f_0(1500)$, a scenario reminiscent of 4 pion spectra with 4 or 2 charged pions in pp CEP in WA experiments [73-76].

The high statistics measurement of $K_sK_s$ pair production in γγ interactions at BELLE [77] confirms the absence of $f_0(1370)$ signal in this process. As can be seen in fig. 12 of ref. [77] the $K_sK_s$ spectrum features dominant $f_2(1270)$ and $f_2'(1525)$ signals and a minimum near 1.4 GeV.

No $f_0(1370)$ signal is present in the $K_sK_s$ spectrum of the charge exchange reaction $K^-p \rightarrow K_sK_s\Lambda$ measured by the LASS spectrometer (see fig. 12 of ref [78]. The spectrum is dominated by the $f_2'(1525)$ signal, and only a small signal from $f_2(1270)$ is present below 1.5 GeV.

## 5. Conclusions and prospects

Independently of the interpretation of the structure in the 1.2-1.5 GeV energy region, quite noticeable is the fact that the STAR $\pi^+\pi^-$ spectrum of fig.1 drops nearly to zero at 1 GeV. This may be the result of the interference of the amplitudes of the low energy tail of the $f_0(1370)$ with the high energy part of the $f_0(980)$ plus the effect of the KKbar threshold, but very likely it might be due to the vanishing of the S.wave continuum for events selected in the low ltl kinematical region. The S-wave continuum, which is usually invoked with its destructive interference with the $f_0(980)$ amplitude to generate the drop at 1 GeV, is drastically reduced in comparison to the AFS data. It looks like the σ meson, which generates the broad structure above 0.5 GeV, is confined below 1 GeV. In other words, under this hypothesis the "red dragon" glueball proposed by Minkowski and Ochs [79], which features a low energy body centered at about 0.6 GeV and a head extending below the $f_0(1500)$, would be split into two separate parts, the σ and the relatively narrow $f_0(1370)$.

The clear confined STAR signal of $f_0(1370)$ decays into $\pi^+\pi^-$ pairs confirms features of $f_0(1370)$ decays observed in pbar annihilations at rest and shows that the σ meson and $f_0(1370)$ are distinct separate objects. The limited width of $f_0(1370)$ and its mass centered around 1370 MeV validate the analysis of data of pbar annihilations at rest into 3 pseudoscalars that used the hypothesis of the possible existence of $f_0(1370)$ as an individual object [42,51,61,57,59,80].

The ratios of the decay branching ratios of $f_0(1370)$ into σσ, ρρ, ππ, KKbar and ηη to the decay branching ratios of $f_0(1370)$ into ππ are of the order of 5.6, 3, 1, 1 and 0.02.

The ratios of the decay branching ratios of $f_0(1500)$ into σσ, ρρ, ππ, KKbar, ηη and ηη' to the decay branching ratios of $f_0(1500)$ into ππ are of the order 0.42, 0.20, 1, 0.22, 0.08, 0.06.

The decay properties of $f_0(1370)$ measured in pbar annihilations at rest and its production properties match the characteristics expected from an object that has a large gluon content [8-10,63].

There is however conflict between the above decay branching ratios of $f_0(1370)$ and the values obtained by WA102 in CEP measurements at 29 GeV [30-32]. The assessment of the values of the branching ratios of decays of the $f_0(1370)$ scalar meson into σσ and ρρ pairs and into ππ, KKbar, ηη pairs of pseudoscalar mesons derived from pbar annihilations at rest is mandatory to establish the nature of $f_0(1370)$.

The STAR data confirm indications of earlier CEP experiments at lower energies and motivates the expectation that moving to higher energies and lower ltl windows will single out pomeron-pomeron interactions, suppress $2^{++}$ production and produce spectra with all low mass $0^{++}$ mesons appearing with little background [19,20,29]. CEP experiments at LHC at low ltl look then extremely promising. The very large energy selects dominant pomeron-pomeron production. Low ltl would select $0^{++}$ production within pomeron-pomeron production. Experiments equipped with precision detectors inside roman pots can approach the circulating beams and measure events with low ltl at both proton vertices. This is the case of the ALFA-ATLAS and CMS-TOTEM Collaborations, that have taken CEP data in 2015 with $\beta^* = 90$ m optics of LHC and a ltl coverage which could go down to 0.03 GeV$^2$ [81]. With CEP data at LHC it should be possible in the medium term to establish with confidence the relative decay branching ratios of $f_0(1370)$ and $f_0(1500)$ (and also of $f_0(1710)$ into $\pi^+\pi^-$, K$^+$K$^-$, K$_s$K$_s$ and compare with the values measured in antiproton annihilations at rest. Other important tasks will be to study of the narrow signal at 1450 MeV observed in CEP experiments at the Omega spectrometer in the $\pi^+\pi^-\pi^+\pi^-$ and $\pi^+\pi^-2\pi^0$ decay channels [73-76] and interpreted as possibly due to the interfering amplitudes of $f_0(1370)$ and $f_0(1500)$ [75,76], to study the other 4 pion decay channels, and to compare the σσ and ρρ decays of $f_0(1370)$ and $f_0(1500)$. In the longer term, in CEP measurements with central detectors tuned to measure both low energy charged and neutral prongs, with larger $\beta^*$ the ltl window could be extended down to 0.003 GeV$^2$ and CEP could be measured in kinematical regimes where $0^{++}$ production should be definitively overwhelming and permit to do a complete study of all low energy scalars.